\documentclass{ws-rv9x6}
\usepackage{subfigure}   
\usepackage{ws-rv-thm}   
\usepackage{ws-rv-van}   
\usepackage{ws-index}    
\usepackage{easybmat}    
\usepackage[version=3]{mhchem}
\makeindex
\newindex{aindx}{adx}{and}{Author Index}       
\renewindex{default}{idx}{ind}{Subject Index}  
\newcommand{\1}{{\rm 1\hspace*{-0.4ex}
\rule{0.1ex}{1.52ex}\hspace*{0.2ex}}}

\begin{document}

\chapter[Diffusion to capture by many competing boundaries]{Diffusion to capture and the concept of diffusive interactions}

\author[F. Piazza]{Marta Galanti$^1$, 
                   Duccio Fanelli$^2$, 
                   Sergey D. Traytak$^3$, 
                   Francesco Piazza$^4$\footnote{Francesco.Piazza@cnrs-orleans.fr}
}

\address{
$^1$Department of Environmental Health Sciences, Mailman School of Public Health, 
Columbia University, 722 West 168thStreet, New York, NY 10032\\
$^2$2 Dipartimento di Fisica e Astronomia, Universit\`a di Firenze, 
INFN and CSDC, Via Sansone 1, 50019 Sesto Fiorentino, Firenze, Italy\\
$^3$Semenov Institute of Chemical Physics RAS, 4 Kosygina St., 117977 Moscow, Russia\\
$^4$Universit\'e d'Orl\'eans, Centre de Biophysique Mol\'eculaire (CBM), CNRS UPR4301,
Rue C. Sadron, 45071, France
}

\begin{abstract}
Diffusion to capture is an ubiquitous phenomenon in many fields in biology and physical
chemistry, with implications as diverse as ligand-receptor binding on eukaryotic and bacterial cells, 
nutrient uptake by colonies of unicellular organisms and the functioning of complex core-shell 
nanoreactors. Whenever many boundaries compete  
for the same diffusing molecules, they inevitably shield a variable part of the molecular flux 
from each other. This gives rise to the so-called {\em diffusive interactions (DI)}, which 
can reduce substantially the influx to a collection of reactive boundaries depending chiefly 
on their geometrical configuration.\\
\indent In this review we provide a pedagogical discussion of the main mathematical aspects 
underlying a rigorous account of DIs. Starting from a striking and deep result on the mean-field 
description of ligand binding to a receptor-covered cell, we develop little by little a rigorous mathematical 
description of DIs in the stationary case through the use of translational addition theorems 
for spherical harmonics. We provide several enlightening illustrations of this powerful mathematical 
theory, including diffusion to capture to ensembles of reactive boundaries within a spherical cavity.

\end{abstract}
\body
\newpage
\tableofcontents

\section{Introduction}\label{s:intro}

It is common knowledge that the first step of any chemical or biochemical 
reaction proceeding in an inert fluid 
phase is the mutual diffusive encounter of reactants~\cite{Rice:1985aa}
\begin{equation}
\label{e:reactionred}
\ce{A + B <=>[\kappa_{\rm on}(t)][k_{\rm off}] A\cdot B <=> ...  <=> P}
\end{equation}
In the above general scheme, the encounter complex $A\cdot B$ is transformed reversibly into 
a (series of) products (collectively indicated by $P$), through a variable number of 
intermediate steps that depend on the specific reaction considered. 
Irrespectively of whether the reaction~(\ref{e:reactionred}) is considered under thermodynamical equilibrium 
or non-equilibrium conditions, the second-order rate constant
$\kappa_{\rm on}$ is proportional to the effective relative diffusion coefficient of the species $A$ and $B$
and describes the formation of the encounter complex. For this reason, these kind of reactions 
are also known in the physical chemistry community with the term {\em diffusion-influenced reactions}.
The time dependence of $\kappa_{\rm on}$ refers to the possibility of investigating transient kinetics effects
as opposed to steady-state (equilibrium or non-equilibrium) kinetics.\\
\indent Several mathematical difficulties emerge when one tries to develop a general kinetic 
theory of  bulk irreversible diffusion-influenced reactions. One way around this problem is to 
treat {\em trapping} and {\em target} models~\cite{Piazza:2013aa}, which are  simpler than the original one. 
In the first model a particle, say, $B$ diffuses towards many sinks $A$ that are 
often assumed to be static. Conversely, the target problem describes a situation where 
many sinks $A$ diffuse to a static particle $B$~\cite{Torquato:2002kx}. 
The two settings, even if equivalent under certain conditions (see Ref.~\refcite{Piazza:2013aa}),
describe in general different problems at high density of reactants.
In the present paper, we shall concentrate on the trapping model for 
reactions of the kind~(\ref{e:reactionred}).\\
\indent Diffusion in domains with one-connected smooth boundaries 
is a fairly well-understood and well-characterized 
phenomenon from the general standpoint of mathematical physics~\cite{Crank:1975aa}.
However, real-life situations are often very complex and make mathematical descriptions challenging. For example, this 
is the case of biochemical reactions taking place in living media, such as the cell interior or the 
extra-cellular matrix (e.g. paracrine delivery)~\cite{Labowsky:2012aa}, where confinement, 
crowding (excluded-volume) effects~\cite{Zhou:2008vf,Verkman:2002ys} and non-specific interactions 
among diffusing species and with all sorts of cellular structures~\cite{Luby-Phelps:2000zr} 
make it very difficult to elaborate quantitative models for the calculation of diffusive encounter 
rates~\cite{Berezhkovskii:2016aa,Traytak:2013aa,Piazza:2013fk,Dorsaz:2010aa,Echeveria:2007aa}.\\
\indent Among all the possible effects on diffusion-influenced encounters and reactions arising 
in non-ideal conditions, in this short review we shall concentrate on the so-called 
{\em diffusive interactions (DI)}. As we shall see in the following, these describe a fundamental 
mechanism of competition among different  reactive boundaries, competing for the same 
diffusive molecular flux. This effect was discussed for the first time 
as far back as 1953 by Frisch and Collins in terms of a {\em competition} 
phenomenon~\cite{Frisch:1953aa}. A decade later, Reck and Prager
referred to the same phenomenon simply as an {\em interaction}~\cite{Reck:1965aa}. 
Although the same problem has been investigated later in different contexts, it seems 
that there is invariably a reference to a generic {\em competition} 
mechanism~\cite{Borzilov:1971aa,Deutch:1976}, until 
the term diffusive interactions was introduced by Traytak~\cite{Traytak:1992}
in analogy with hydrodynamical interactions for Stokes flow in many-body systems.\\
\indent Picture for example a concentration field of small ligand molecules (e.g. signaling hormones or growth factors) diffusing in the extra-cellular matrix or in the bacterial periplasm looking for an available receptor on a cell membrane to form a complex. For the sake of the argument, let us imagine that far away from the receptor-covered
surface the ligand bulk concentration is constant and that the ligand-receptor affinity is large enough 
to consider receptors as {\em sinks}, i.e. perfectly absorbing units. H. Berg has famously termed this general 
scheme of diffusive problems {\em diffusion to capture}~\cite{Berg:1993aa}.
If the surface density of receptors is low, then the diffusion problem is additive and 
the overall capture rate for the whole ensemble of receptors (e.g. number of ligands diffusing to a receptor
per unit time) is well estimated by the sum of the individual rates. In this case, a many-body problem can be solved 
as many identical two-body problems. However, receptors on cell membranes 
are typically very densely packed in clusters~\cite{Sourjik:2004aa,Lelievre:2004aa,DeLisi:1981aa,Briegel:2009aa,Balint:2017aa,Angioletti-Uberti:2017aa,Hlavacek:1999fk}. In this case, any two identical receptors sitting at close 
separation will {\em screen} a portion of the diffusive ligand flux to each other. Overall, a complex pattern
of many-body screening effects will arise, reflecting the many-body geometrical arrangement 
of receptors, with the consequence of reducing the overall capture rate corresponding to the array of receptors. 
A similar scenario is relevant  for the case of multivalent molecules, i.e molecules 
carrying more than one binding sites~\cite{Vauquelin:2013aa,Todorovska:2001aa,Nunez-Prado:2015aa,Fasting:2012aa,Mammen:1998xd}.  
By the same token, diffusive interactions among the different active sites will give rise to
a similar negative cooperativity, which will reduce the overall capture rate with respect to 
an equivalent number of isolated sites.\\
\indent Mathematically, if many-body effects are 
relatively well characterized for unbounded systems of distributed 
sinks~\cite{Gopich:2002uq,Berezhkovskii:1999ua,Yuste:2008aa,Tachiya2007,Piazza:2013aa,Keizer:1987aa,Felderhof:1976mw}, 
the case of diffusive interactions for sinks or partially absorbing boundaries located 
in a finite domain is more challenging from a mathematical standpoint. This scenario is relevant in many 
fields, ranging from  catalysis in composite nanostructures~\cite{Galanti:2016aa,Piazza:2015aa,Lu:2011aa}
to nutrient uptake by dense colonies of microorganisms~\cite{Sozza:2018aa}.\\
\indent A full treatment of time-dependent diffusive interactions is extremely hard to treat and
{\em de facto} limited to simple cases~\cite{Traytak:1995fk,Traytak:2008ys}.
Conversely, several methods have been used to tackle this problem for different geometries in the stationary state,
where more theoretical approaches are available, such as renormalization group~\cite{Traytak:1992,Traytak:1996aa} 
and the {\em method of irreducible Cartesian tensors}~\cite{Traytak:1992} and 
the {\em generalized method of separation of variables}~\cite{Russa:2009,Galanti:2016ab},
based on addition theorems for solid harmonics~\cite{Morse:1953le,Cahola:1978,Russa:2009,Galanti:2016ab}.
It is also possible to combine such methods with methods based on dual-series relations~\cite{sneddon:1966}
to deal with the case of diffusive interactions among inhomogeneous reactive boundaries with 
active and reflecting patches~\cite{Traytak:2007aa,Piazza:2005vn,Traytak:1995}.\\
\indent In this short review, we will provide a concise account of DI arising in many-body systems  
consisting of spherical fully absorbing and partially absorbing 
boundaries within a finite domain. In~\sref{s:1} we will lay out 
the basic ideas of the mathematical method employed to compute the diffusive encounter rate 
for two isolated spherical molecules. In~\sref{s:2}, these ideas will be used to 
estimate many-body effects in the mean-field approximation and illustrate the main physical
features of DI. In~\sref{s:3}, we will show how using multipole expansion methods coupled to 
translational addition theorems for solid harmonics allows one to solve the problem 
exactly to any desired level of accuracy for arbitrary geometries. This method
constitutes a  powerful tool that can be employed to tackle a wide host of 
important problems in physical chemistry and biology.  

\section{Bimolecular diffusive encounters as two-body boundary problems\label{s:1}}

The polish physicist Marian Ritter von Smolan Smoluchowski (1872-1917), besides being 
a skilled water-color painter and an exquisite pianist, during the first two decades of the XX century
laid the bases of the mathematical 
theory of diffusion processes~\cite{Smoluchowski:1916fk,Smoluchowski:1917aa}. The calculation 
of diffusive encounter rates in ideal conditions follows directly from his ideas.
Let us imagine a solution containing two spherical molecules $A$ and $B$, with radii $R_A$ and $R_B$,
diffusion coefficients $D_A$ and $D_B$ and concentrations (number densities) $c_A$ and $c_B$.
Our goal is to determine the rate of $A$-$B$ encounters dictated by their relative diffusive motion as
a function of the concentrations. The full many-body problem
is exceedingly hard to treat analytically. However, as it was
first recognized by Smoluchowski, one can reduce
it to the effective two-body problem of relative diffusion of a
single A-B pair under certain hypotheses. However, as already
noted by Szabo~\cite{szabo1989theory}, the commonly accepted hypothesis of
high dilution of both species is not enough. The first step
towards the equivalent two-body problem is that one species
be much more diluted than the other.  
Yet, not even this is enough.  
Let us imagine that the particles of kind $A$ are
sufficiently diluted, i.e. $c_A\ll c_B$, 
so that one can concentrate on a single $A$
particle surrounded by many $B$ particles, say $N$ of them. 
It is not difficult to show that the $(N + 1)$-body Smoluchowski equation 
describing the diffusion of a single $A$ molecule within a sea of $B$ particles
contains cross-terms that make it non-separable if $D_A \neq 0$~\cite{Piazza:2013aa}.
Therefore, bimolecular encounters between $A$ and $B$ molecules 
can be modeled as an equivalent two-body problem provided that
\begin{romanlist}
\item Both species should be highly diluted, so that mutual interactions
can be safely neglected.
\item One species ($A$) must be much more diluted than the other, so
that the full problem can be reduced to study the fate of a
single particle surrounded by many particles of the other species.
\item The diffusion coefficient of the highly diluted species
should be much smaller than that of the other species
(from $N$-body to two-body). A consequence of this is that 
the relative diffusion coefficient essentially 
coincides with the one of the mobile species, i.e. 
$D = D_A+D_B \simeq D_B$.
\end{romanlist}
\noindent Under these conditions the rate of encounters can be computed by solving the 
following stationary diffusion problem (i.e. Laplace equation) 
for the local concentration field $c(\boldsymbol{r})$ of $B$ molecules.
\begin{subequations}\label{e:SmolBP}
\begin{align}
&\nabla^2 c = 0 \label{e:SmolBP1}\\
&c|_{\partial \Omega} = 0 \label{e:SmolBP2}\\
&\lim_{r\to\infty} c = c_B \label{e:SmolBP3}
\end{align}
\end{subequations}
This describes to the non-equilibrium steady state arising as a constant bulk concentration $c_B$
is maintained far from the  reactive boundary $\partial \Omega$, which acts as a perfectly 
absorbing sinks. This is the contact surface of the two molecules, which is the sphere $\mathcal{S}_R$
of radius $R = R_A + R_B$, since both $A$ and $B$ molecules are spherical by assumption. 
Physically, this describes the pseudo-first order (annihilation) reaction
\[
A+B\stackrel{\kappa_S}{\to} A
\]
whose rate $k_S = \kappa_S c_B$ coincides with the overall 
flux into the reactive surface $\mathcal{S}_R$, i.e. the number 
of $B$ molecules crossing $\mathcal{S}_R$ per unit time. Here, we shall use the Greek letter $\kappa$ to 
denote rate constants (dimensions of inverse concentration times inverse time) and the Latin 
letter $k$ to denote rates (dimensions of inverse time). 
The rate can be computed straightforwardly as the incoming flux across $\mathcal{S}_R$, that is 
\begin{equation}
\label{e:flux}
k_S =  -\int_{\mathcal{S}_R} \boldsymbol{J}\cdot \hat{\boldsymbol{n}} \, dS
\end{equation}
where $ \boldsymbol{J} = -D\nabla c$ is the relative diffusion current (Fick's first law). \\
\indent The solution to the spherically symmetric boundary problem~(\ref{e:SmolBP}) 
can be computed straightforwardly, yielding
\begin{equation}
\label{e:Smolrho}
c(r) = c_B \left(  1 - \frac{R}{r} \right)
\end{equation}
which, using~\eref{e:flux}, immediately gives the so-called Smoluchowski rate constant $\kappa_S$
\begin{equation}
\label{e:Smolk}
\kappa_S = 4\pi D R
\end{equation}
The rate of encounter is thus $k_S =  4\pi D R c_B$ molecules per unit time  disappearing 
across the absorbing boundary $\mathcal{S}_R$. Note that this is proportional to the {\em linear}
size of the latter, a distinctive signature of the diffusive dynamics.\\
%
%
\subsection{Finite reaction probability and radiation boundary conditions}

The scheme~(\ref{e:SmolBP}) describes the reactive boundary as a perfect sink, which 
amounts to consider that $B$ particles are annihilated the moment they reach 
contact distance. This is meant to describe a (long-lived) binding event. 
However, in reality the two reacting partners first approach diffusively to contact distance 
forming the so-called {\em encounter complex}. Subsequently, this can either dissociate (this is 
the case if for example the two partners were not mutually oriented in a favourable manner) 
or proceed to form a stable contact. This more realistic situation can be accommodated for
within the above mathematical formalism thanks to an intuition put forward by Collins and Kimball 
in 1949~\cite{collins1949diffusion}. The idea is to replace the perfectly absorbing boundary 
condition~(\ref{e:SmolBP2}) with a {\em radiation} boundary condition (in more 
mathematical terms Robin boundary condition),
that interpolates between  perfectly absorbing (Dirichlet type) and perfectly reflecting (von Neumann type)
boundary conditions. Namely, the boundary value problem~(\ref{e:SmolBP}) is replaced by 
\begin{subequations}\label{e:SmolBPrbc}
\begin{align}
&\nabla^2 c = 0 \label{e:SmolBPrbc1}\\
&\left.\left( 
    4\pi D R^2 \frac{\partial c}{\partial r} - \kappa^\ast c
 \right)\right|_{\partial \Omega} = 0 \label{e:SmolBPrbc2}\\
&\lim_{r\to\infty} c = c_B \label{e:SmolBPrbc3}
\end{align}
\end{subequations}
The boundary condition~(\ref{e:SmolBPrbc2}) stipulates that the particle flux across the reactive 
contact surface is proportional to the local concentration of ligands $B$. The proportionality 
constant, the {\em intrinsic} rate constant $\kappa^\ast$, can be considered as describing the physical mechanism
underlying the chemical fixation of the encounter complex. In the limit $\kappa^\ast\to 0$, the 
surface becomes perfectly reflecting, i.e. no reaction can occur. In the opposite limit,
$\kappa^\ast\to \infty$ (mathematically, it is necessary to divide~\eref{e:SmolBPrbc2} by $\kappa^\ast$ before taking the 
limit) one recovers the perfectly absorbing boundary, which is thus seen as corresponding to infinitely 
fast chemical fixation step. As we shall see, the latter case is known as the {\em diffusion-limited} regime,
where the diffusive encounter is the rate-limiting step of the reaction. Diffusion-limited is {\em as-fast-as-one-can-go},
other situations corresponding to a finite intrinsic reaction rate necessarily proceeding slower than that. \\
\indent The solution to the problem~(\ref{e:SmolBPrbc}) can be computed as straightforwardly as before, yielding
\begin{equation}
\label{e:Smolrho}
c(r) = c_B \left[  1 - \left(  \frac{h}{1+h}\right)\frac{R}{r} \right]
\end{equation}
where $h = \kappa^\ast/\kappa_S$,
which, using~\eref{e:flux},  gives 
\begin{equation}
\label{e:Smolk}
\kappa = \kappa_S \left(  \frac{h}{1+h}\right)
\end{equation}
In general, $\kappa < \kappa_S$. This case in the physical chemistry community is often indicated 
with the specific term {\em diffusion-influenced} regime (or reaction), as opposed to the  
diffusion-limited regime, $h\to\infty$, where $\kappa = \kappa_S$.

\section{Approximate evaluation of diffusive interactions: a surprising lesson in biology\label{s:2}}

The concept of diffusive interactions is best introduced through a simple, yet astonishing classical 
result. Let us consider a cell, which we model as a spherical surface of radius $R$, uniformly covered 
with $M$ receptors, which we model as small absorbing circular patches of radius $a$. The rest
of the cell surface is supposed to be reflecting. The problem of computing the rate of absorption 
of this partially absorbing cell was first famously considered by Berg and Purcell in 1977~\cite{Berg:1977aa}.
Here, we shall follow the appealing re-derivation by Shoup and Szabo~\cite{Shoup:aa} of the 
same result. The main idea is to treat the receptor-covered cell as a partially absorbing sphere, 
in the sense of radiation boundary conditions. According to Shoup and Szabo's argument, the corresponding 
intrinsic reaction rate constant can be computed as the ratio between the rate constant of $M$
isolated circular disks on an otherwise reflecting surface and the Smoluchowski rate 
constant of the entire cell, namely
\begin{equation}
\label{e:SSkstar}
\kappa^\ast =  4Da \times M
\end{equation}
where we used the classical result $k_a = 4Da$ for the rate constant of a small absorbing disk 
on an infinite reflecting plane~\cite{Hill:1975aa}.
Using~\eref{e:Smolk},  the rate constant corresponding to the partially 
absorbing sphere is easily found, namely
\begin{equation}
\label{e:SSkcell}
\kappa = \kappa_S \left( \frac{Ma}{\pi R + Ma} \right)
\end{equation}
A surprising finding emerges if we plug realistic figures in~\eref{e:SSkcell}.
The typical size of a cell is around $10$ $\mu$m, while the typical size of a receptor
is of the order of $1.5$ nm. If we calculate how many receptors are needed to reduce the rate
constant to only one half that of the fully covered cell, i.e. $\kappa_S$, we find 
$M \simeq 10^4$, which is the correct order of magnitude for the average number 
of receptors of a given family present on a cell's surface at any given time~\cite{Lauffenburger:1993aa}.
Is this a large number? A quick calculation shows that  
the fraction of cell surface covered by as many receptors is $\simeq 10^{-4}$ !
To summarize, an active surface fraction as low as $10^{-4}$ only yields a 
factor of 2 reduction in the rate of capture. The surprising finding is that 
an extremely sparse uniform distribution of receptors is as effective an absorber 
as a fully covered cell. \\
\indent We can now ask a deep and intriguing question. What would be the rate
if all the $M$ receptors were clustered in one single active patch covering the 
same surface fraction? The answer to this question is the result of  
the classical calculation of the rate to an active spherical cap on an otherwise 
reflecting sphere~\cite{Traytak:1995}, $\kappa = f_c(\theta_0) \kappa_S$,
where $f_c(\theta_0)\leq 1$ is a steric factor 
describing the diffusion to the active cap of aperture $\theta_0$.
In the monopole approximation (MOA), one has
\begin{equation}
\label{e:stericf}
f_c(\theta_0) = \frac{\sin \theta_0 + \theta_0}{2\pi - (\sin \theta_0 + \theta_0)}
\end{equation}
Incidentally, according to the general physics of diffusion, $f_c$ can be approximated as 
the {\em square root} of the surface fraction covered by the cap~\cite{Traytak:1995}, namely 
\begin{equation}
\label{e:stericfapp}
f_c(\theta_0) \approx \sqrt{\frac{\Delta S_{\rm cap}(\theta_0)}{4\pi R^2}} = 
          \sqrt{M}\left( 
                    \frac{a}{2R}
                  \right)
\end{equation}
Combining~\eref{e:SSkcell} and~\eref{e:stericfapp}, we can compute the ratio between the steric factor $f_u$
corresponding to a sparse uniform configuration  of the $M$ receptors and that of the cluster configuration, $f_c$,
namely
\begin{equation}
\label{e:sterred}
\frac{f_{u}}{f_c} \simeq \frac{2\sqrt{M}}{\pi} + \mathcal{O}\left(a/R\right)
\end{equation}
For a number of receptors $M$ of the order of  $10^4\div10^5$, one finds $f_u/f_c \simeq 10^2$.
This is a first, striking manifestation of the anticooperative effects caused by diffusive interactions.
Summarizing, (i) the rate of capture for a ligand diffusing to a cell uniformly and very sparsely
covered with receptors is essentially as large as that of a fully covered cell and (ii) about 100 times 
larger than in the case where all the receptors would be clustered in a single active patch. 
It is intriguing to observe than in many cases receptors are indeed densely clustered 
on the cell surface. Famously, this is the case of chemotaxis receptors in bacteria such as 
E. Coli, forming extended patches at the cell poles~\cite{Sourjik:2004aa,Briegel:2009aa,Balint:2017aa}.
One might argue that there should be other biochemical or structural constraints that offset such strong 
reduction to the rate of capture.

\section{The generalized method of separation of variables allows one to solve the problem 
semi-analytically\label{s:3}}

A precursor idea of the {\em generalized method of separation of variables} (GMSV), first discussed
in 1944 by S. K. Mitra~\cite{Mitra:1944aa} relating to Laplace equation with two 
disconnected spherical boundaries, goes back to the  well-known paper by Lord Rayleigh 
on the conductivity of heat and electricity in a medium with regularly 
arranged obstacles~\cite{Rayleigh:1892aa}.\\
\indent In the theory of partial differential equations (PDE),
a 3D (bounded or unbounded) domain $\Omega \subset \Bbb{R}^3 $ is called a 
{\em canonical domain} for a given PDE if the classical 
solution to this equation may be expanded in an absolutely and uniformly 
convergent series with respect to corresponding basis solutions in the Hilbert 
space $L_2 (\partial \Omega)$.
Remarkably, the GMSV allows one to find  semi-analytical solutions of 
various boundary value problems for Laplace equation in all known 3D canonical 
domains and their combinations thereof~\cite{Traytak:2018aa}.
The GMSV can be thought of comprising five separate logical steps, 
\begin{alphlist}
\item reduction of the boundary value problem to its non-dimensional standard form
\item determination of the basis solutions to the equation in a given canonical domain
\item application of the linear superposition principle
\item application of the re-expansion (addition) theorems in order to impose the boundary conditions
\item reduction of the problem to an infinite system of linear algebraic equations and its solution.
\end{alphlist}
\indent In principle, one would like to solve the problem of diffusion to ensembles of absorbing or 
partially absorbing boundaries exactly. Although  the GMSV  
can be used to deal with (general) canonical domains, we will limit ourselves here to only spherical boundaries.

\subsection{Diffusive interactions between two spheres}

\begin{figure}[t!]
\centerline{\includegraphics[width=10cm]{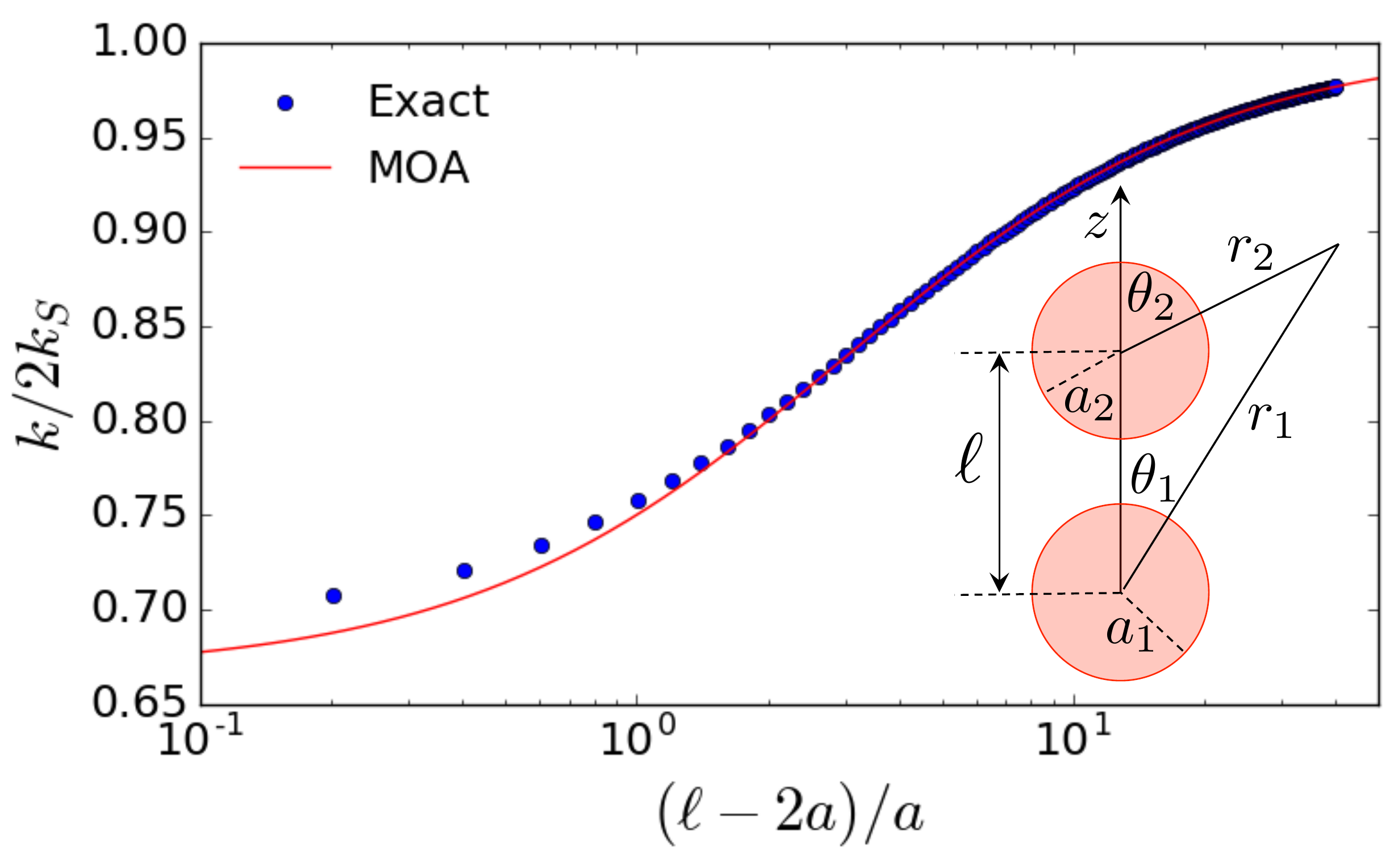}}
\caption{Rate of capture for two sinks of radius $a_1=a_2=a$ separated by a distance $\ell$.
Comparison of the exact result and the MOA approximation.}
\label{f:twosinksMOA}
\end{figure}

The general power of the GMSV and its main features can be most 
clearly appreciated by discussing a simple problem, 
namely that of diffusion to a pair of spherical sinks of radius $a_1$ and $a_2$
located  at the origin ($\Omega_1$) and along the $z$ axis at $z=\ell$, ($\Omega_2$). 
The diffusion of ligands (particles $B$) should be described in  
the 3D smooth oriented manifold $ \Omega^{-}=\Bbb{R}^3\backslash \overline{\Omega }_1\cup\overline{\Omega }_2$,
which can be referred to as the {\em concentration manifold}~\footnote{It is expedient to introduce 
also the {\em partial domains} $\mathcal{D}_i = \Bbb{R}^3\backslash \overline{\Omega }_i $, 
so that $\Omega^{-}=\mathcal{D}_1\cap \mathcal{D}_2$.}.
With reference to the logical sequence of the GMSV, we proceed as follows. 
\paragraph{(a)} To find the standard non-dimensional form of the problem, we consider the  reduced 
concentration field of $B$ particles that is regular at infinity, that is, 
\[
u(\boldsymbol{r}) = 1 - u(\boldsymbol{r})/c_B
\]
and non-dimensional radial coordinates $\xi_i = r_i/a_i$. The non-dimensional 
standard form of the original boundary value problem reads
\begin{subequations}\label{e:2sph}
\begin{align}
&\nabla^2 u = 0 \label{e:2sph1} \quad \mbox{in} \quad \Omega^{-} \\
&u|_{\xi_i = 1} = 1 \label{e:2sph2}\\
&u|_{\xi_i \rightarrow \infty} \rightarrow 0  \label{e:2sph4}
\end{align}
\end{subequations}
\paragraph{(b)} The appropriate basis functions for this problem are  
scalar axially symmetric {\em regular} and {\em irregular solid spherical harmonics}
with respect to the two spherical coordinate systems for $\Omega_i$ (see cartoon 
in~\fref{f:twosinksMOA})  
\begin{equation}
\psi _{n}^{+}(r_i,\theta_i )=r_i^n\,P_{n}(\mu_i), \qquad  
\psi _{n}^{-}(r_i,\theta_i)={r_i^{-n-1}}P_{n}(\mu_i),   \label{green6a}
\end{equation}
where $P_{n}(\mu_i)$ is a Legendre polynomial of degree $n$, with 
$\mu_i=\cos\theta_i$. Solid spherical harmonics form a {\em canonical basis}, 
$ \lbrace \psi _{n}^{+}(r_i, \theta_i) \rbrace_{n=0}^{\infty} $ and 
$ \lbrace \psi _{n}^{-}(r_i, \theta_i) \rbrace_{n=0}^{\infty} $, 
for harmonic functions in $\Omega_i $ and $\mathcal{D}_i$, respectively.
\paragraph{(c)} For $N>2$ it is impossible to introduce a global coordinate system (e.g.
bispherical coordinates for $N=2$, such as in ref.~\refcite{Samson:1977aa}). 
Hence, in general one should introduce appropriate local coordinates in $\Omega^{-}$.
The solution to the problem~(\ref{e:2sph}) can be expressed as 
\begin{equation}
\label{e:2sphu1}
u(\boldsymbol{r})= u_1(\boldsymbol{r}_1) + u_2(\boldsymbol{r}_2)\quad \mbox{for}\quad\boldsymbol{r}_i\in 
\mathcal{D}_i  
\end{equation}
where 
\begin{equation}
\label{e:2sphu2}
u_i(\boldsymbol{r}_i) = \sum_{n=0}^\infty A_n^i \psi _{n}^{-}(\xi_i, \theta_i) \quad \mbox{in}\quad \mathcal{D}_i
\end{equation}
are absolutely and uniformly convergent series expansions of irregular spherical harmonics.
The unknown coefficients $A_n^i$ should be determined by imposing 
the boundary conditions~(\ref{e:2sph2}) for $ i=1,2 $. In order to do so, we have to express the function
$u_1(\boldsymbol{r}_1)$ in the local coordinates of $\Omega_2$ and viceversa. 
\paragraph{(d)} This can be accomplished through {\em addition theorems}~\cite{Morse:1953le}.
For the present axially symmetric problem, one has
\begin{eqnarray}
\label{e:ATas}
\xi_1^{-(k+1)}P_k(\mu_1) = \sum_{n=0}^\infty U^{21}_{nk}   \,\xi_2^n  P_n(\mu_2)\qquad \mbox{for}\quad  \xi_2<\ell \label{e:ATas1}\\
\xi_2^{-(k+1)}P_k(\mu_2) = \sum_{n=0}^\infty U^{12}_{nk} \,\xi_1^n  P_n(\mu_1)\qquad \mbox{for}\quad  \xi_1<\ell \label{e:ATas2}
\end{eqnarray}
where $\epsilon_i = a_i/\ell < 1$ and the so-called {\em mixed-basis matrices elements} read
\begin{subequations}
\label{e:U12mat}
\begin{align}
&U^{12}_{nk} = (-1)^k \binom{n+k}{n} 
                      \epsilon_1^{n}\epsilon_2^{k+1} \label{e:U12mat1}\\
&U^{21}_{nk} = (-1)^n \binom{n+k}{n} 
                      \epsilon_1^{k+1}\epsilon_2^{n} \label{e:U12mat2}
\end{align}
\end{subequations}
\Eref{e:ATas2} needs to be used when imposing that $u(\boldsymbol{r})$ satisfy~\eref{e:2sph2} for 
$i=1 $ and~\eref{e:ATas1} needs to be used for $i=2$. 
\paragraph{(e)} This procedure leads to the following  
infinite system of linear algebraic equations of the II kind (ISLAE),
comprising in general as many equations as there are boundaries,
\begin{equation}
\label{e:2sphLS}
\left\{
\begin{array}{l}
A^1_n+\sum_{k=0}^\infty U^{12}_{nk} A_k^2 = \delta_{n0}   \\[10pt]
\sum_{k=0}^\infty U^{21}_{nk} A_k^1 + A_n^2 = \delta_{n0} 
\end{array}
\right.
\end{equation}
It may be proved that the system~(\ref{e:2sphLS}) can be truncated to obtain 
a solution to any desired accuracy through the so-called {\em reduction
method}~\cite{Kantorovich:1982aa}.\\
\indent The overall rate of capture $k$, i.e. the total flux into the two-sphere system, 
is given by 
\begin{eqnarray}
\label{e:k2sph}
k &=& -2\pi D a_1 \int_{-1}^1 \left. \frac{\partial u}{\partial \xi_1} \right|_{\xi_1=1}  \, d\mu_1
      -2\pi D a_2 \int_{-1}^1 \left. \frac{\partial u}{\partial \xi_2} \right|_{\xi_2=1}  \, d\mu_2 \nonumber\\
  &=& k_{S_1} A^1_0 + k_{S_2} A^2_0     
\end{eqnarray}
where we have used the general property of Legendre polynomials $\int_{-1}^1 P_n(\mu)\,d\mu = 2\delta_{n0}$
and introduced the two Smoluchowski rates, $k_{S_i} = 4\pi D a_i c_B$. \\
\indent The simplest analytical approximation of the exact solution is the {\em monopole} approximation (MOA), which 
consists in keeping only the $n=0,k=0$ terms in the system~(\ref{e:2sphLS}). It is not difficult to see that this yields
\begin{equation}
\label{e:k2sphMOA}
k = k_{S_1} \left( 
              \frac{1-\epsilon_2}{1-\epsilon_1\epsilon_2}
            \right) + 
    k_{S_2} \left( 
              \frac{1-\epsilon_1}{1-\epsilon_1\epsilon_2}
            \right)  \leq  k_{S_1} + k_{S_2}     
\end{equation}
The case of two equal sinks provides some immediate insight into the anticooperativity of 
diffusive interactions. If $a_1=a_2$,~\eref{e:k2sphMOA} reduces 
to the well-known result~\cite{Deutch:1976}
\begin{equation}
\label{e:k2sphiMOA}
k = \frac{2k_S}{1+\epsilon} =  2k_S\frac{\ell}{a + \ell}  
\end{equation}
%
%
It can be appreciated that $k\to 2k_S$ in the limit of infinite separation, $\ell\to\infty$.
The MOA predicts a maximum reduction $k/2k_S=2/3$ of the rate of capture (i.e. maximum strength of DIs) 
at contact distance, $\ell=2a$.  
This has to be compared with the exact value~\cite{Samson:1977aa}, $k/2k_S = \log 2\approx 0.693$.
It is interesting to observe that DIs are long-range, that is $1- k/2k_S \simeq a/\ell$ for separations 
larger than a few radii: DIs are entropic forces that decay with distance 
like Coulomb and gravitational interactions.\\
\indent  One may wonder how good an approximation is the MOA.  It turns out that for assemblies of 
perfectly absorbing sinks it is indeed an extremely good approximation, as it is apparent 
from~\fref{f:twosinksMOA}. For $\ell=3a$ the relative error is less than 1 \%. It can be shown that 
the relative error decreases rapidly, $\propto \ell^{-4}$, until $\ell \simeq 10 a$ (approximately 0.01 \%), 
and then decreases more slowly, $\propto \ell^{-1}$. 
The reasons why the MOA is so good an approximation have been investigated in 
Ref.~\refcite{Traytak:1992}.

\subsection{Diffusive interactions are weaker among multiple partially reactive boundaries\label{ss:3.1}}

A system of partially reactive boundaries experiences weaker diffusive interactions. This can be easily 
seen and quantified by repeating  the above calculations for two spheres endowed with 
intrinsic rate constants $\kappa_1$ and $\kappa_2$. This entails replacing boundary conditions~(\ref{e:2sph2}) 
with radiation conditions, namely
\begin{equation}
\label{e:2sphh}
\left[ \frac{\partial u}{\partial \xi_i}  - h_i (u-1) \right]_{\xi_i=1} = 0
\end{equation}
where $h_i = \kappa_i^\ast/\kappa_{S_i}$, $i=1,2$. 
In this case, it is not difficult to take the same steps as in the above derivation 
and compute the new matrices $U^{12}$, $U^{21}$. The MOA gives in this case
\begin{equation}
\label{e:k2sphhMOA}
k = k_{S_1} \left[
              \frac{q_1(1-q_2\epsilon_2)}{1-q_1q_2\epsilon_1\epsilon_2}
            \right] + 
    k_{S_2} \left[ 
              \frac{q_2(1-q_1\epsilon_1)}{1-q_1q_2\epsilon_1\epsilon_2}
            \right]       
\end{equation}
where $q_i = h_i/(1+h_i)$. The case of two identical, partially absorbing spheres gives immediately
\begin{equation}
\label{e:k2sphidhMOA}
\frac{k}{2k_{S}} = \frac{h}{1+h(1+\epsilon)} \leq \frac{1}{1+\epsilon}
\end{equation}
Diffusive interactions are therefore less prominent for partially absorbing boundaries. It is easy to 
check that the maximum strength of DIs (i.e. at contact distance) is reduced by an intrinsic reaction 
rate $\kappa^\ast$ by a factor $3h/(2+3h)$ in the MOA.  However, it should be emphasized that the MOA
performs increasingly worse the lower the value of $h$, and more multipoles 
should be considered beyond the $n=0$ term to achieve the same accuracy as in the limit $h\to \infty$. 
%

\section{Many spherical boundaries arranged arbitrarily in space\label{s:4}}

\begin{figure}[t!]
\centerline{\includegraphics[width=7cm]{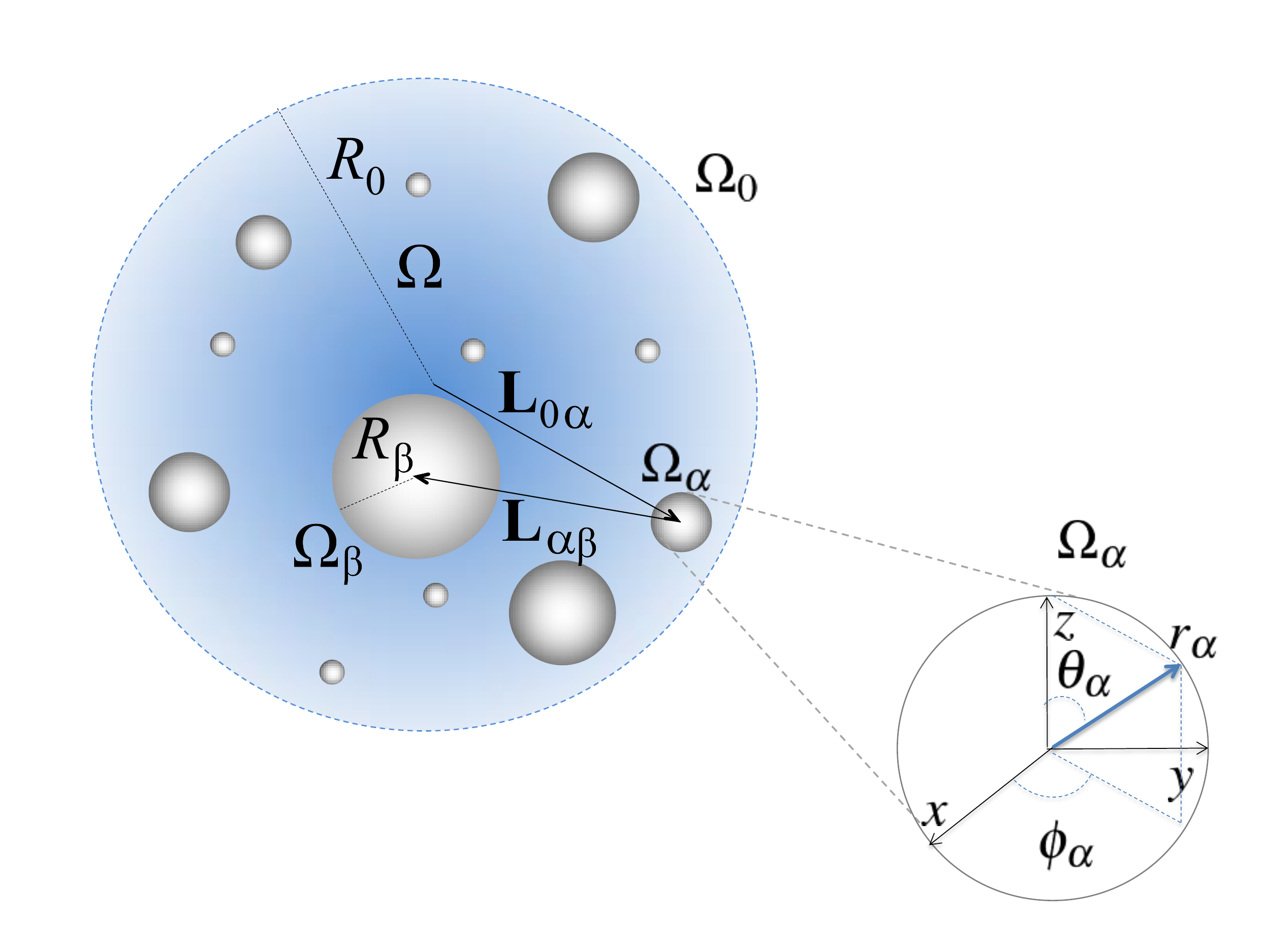}}
\caption{Schematic representation of the domain $\Omega$ with the relevant local coordinate systems,
radii and position vectors.}
\label{f:schemacoord}
\end{figure}

The trick of using addition theorems to express multipole expansions in local reference frames 
centered on two different disconnected spherical boundaries can be extended with no 
conceptual difficulties to the case of many spheres of arbitrary size, intrinsic reaction rate 
constant and position in 3D space. 
Let us consider the finite spherical domain 
$\displaystyle \Omega=\Omega_0\setminus \bigcup_{\alpha=1}^{N}\overline{\Omega}_\alpha$,
represented in~\fref{f:schemacoord}, filled with $N$ spherical reactive boundaries.
Let us introduce the non-dimensional normalized ligand density $ u(\bold{r})=c(\bold{r})/c_B$ 
and the variables $\xi_\alpha=r_\alpha/R_\alpha$, $\xi_0=r_0/R_0$, normalized to the radii of the 
respective reactive boundaries. We need to solve the following boundary problem
\begin{subequations}
\label{sistema_def}
\begin{align}
&\displaystyle\bigtriangledown^2 u=0  \label{sistema_def1}\\
&\displaystyle 
\bigg( \left.\frac{\partial u}{\partial \bold{\xi_\alpha}}-h_\alpha u\bigg)\right|_{\partial \Omega_\alpha}=0\qquad \forall \, \alpha=1,2,\dots,N \label{sistema_def2}\\
&\displaystyle \left.\bigg(\frac{\partial u}{\partial {\xi_0}} + 
h_0(u - 1)\bigg)\right|_{\partial \Omega_0}=0 \label{sistema_def3}
\end{align}
\end{subequations}
Again, we have introduced the parameters $h_\alpha={\kappa_\alpha^\ast}/{4 \pi D R_\alpha}$ that determine 
the reactivity of the $\alpha$-th sphere. The boundary condition~(\ref{sistema_def3}) on the 
inner surface of the {\em container} sphere $\Omega_0$ is a radiation-type boundary condition
and has the following meaning. One should imagine that the ligand concentration is $c_B$ outside 
$\Omega_0$ (even if formally the problem is not defined there) and that there is a membrane 
separating the inner compartment $\Omega$ from the exterior whose non-dimensional permeability
is proportional to $h_0$. In the limit $R_0\to\infty$, one recovers the open-boundary problem with 
the boundary condition $\lim_{R_0\to\infty} c=c_B$. Furthermore, it is not difficult to show
that if one considers the problem~(\ref{sistema_def}) for a single sink at the center of $\Omega_0$
and an equivalent problem (single sink) in the open domain but with 
$D = \{D_{\rm in} \ \text{for} \ r\leq R_0 | \ D_{\rm out} \ \text{for} \ r> R_0\}$, then the two 
problems are equivalent provided $h_0 = D_{\rm out}/D_{\rm in}$. Hence, one may think of 
the problem~(\ref{sistema_def}) as describing diffusion of  ligand to a set of spheres within a spherical 
container such that the ligand concentration outside the container is fixed ($c_B$), as well as 
the ratio $h_0$ between the ligand diffusion coefficient outside the container (bulk) 
and in the interior, the latter parameter playing the role of the non-dimensional permeability 
of the (imaginary) membrane at $\partial \Omega_0$. \\
\indent By virtue of the superposition principle for the Laplace equation, 
the problem~(\ref{sistema_def}) admits a solution in $\Omega$ as a sum of linear 
combinations of regular (inside $\Omega_0$) and irregular harmonics 
(outside each $\Omega_\alpha$), namely
\begin{eqnarray}
\label{e:gensolu}
u  &=& u^+_0+\sum_{\alpha=1}^N u^-_\alpha \nonumber \\
   &=& \sum_{n=0}^{\infty}\sum_{m=-n}^{n}A_{mn} \xi_0^nY_{mn}(\boldsymbol{r}_0) + 
       \sum_{\alpha=1}^N\sum_{n=0}^{\infty}\sum_{m=-n}^{n}B_{mn}^\alpha \xi_\alpha^{-n-1}
             Y_{mn}(\boldsymbol{r}_\alpha)
\end{eqnarray}
where $Y_{mn}(\boldsymbol{r}_\alpha) = P^m_n(\cos\theta_\alpha) e^{im\phi_\alpha}$ are 
solid harmonics referring to the local reference frame centered on the $\alpha$-th boundary 
(see~\fref{f:schemacoord}).
The coefficients $A_{mn},B^\alpha_{mn}$ should be determined by imposing the boundary conditions. 
In the neighborhood of each boundary one has to express all the bases as a function of the local coordinates.
More precisely, in the neighborhood of each $\partial \Omega_\alpha$, $u^+_0$ and $u^-_\beta$ ($\beta\neq \alpha$)
have to be expressed as a function of the $\boldsymbol{r}_\alpha$ coordinates, 
and similarly, in the neighborhood of $\partial \Omega_0$, every $u^-_\alpha$  
has to be written as a function of $\boldsymbol{r}_0$. 
For this purpose, one can make use of the translational addition theorems (AT) for solid 
harmonics~\cite{Morse:1953le}. This operation requires some care, as one out 
of three possible ATs must be selected for each pair of boundaries depending 
on the geometry. These rules are summarized in appendix~\ref{a1}.
%
%
\subsection{Many spheres inside a spherical cavity\label{s:5}}
Diffusion-influenced reactions inside a spherical cavity are of great importance in various applications, 
however, often the simple Smoluchowski rate is incorrectly used to describe the kinetics of these 
reactions~\cite{Sneppen:2005aa,Vazquez:2010aa}.\\
\indent The rate of capture of a sink of radius $R_1$ at the center of a spherical cavity of radius $R_0$ 
outside which there is a constant bulk ligand density $c_B$ is given by the solution of the following problem
\begin{subequations}
\label{e:1sinkcav}
\begin{align}
&\displaystyle\bigtriangledown^2 u=0  \label{e:1sinkcav1}\\
&u|_{r=R_1}=0  \label{e:1sinkcav2}\\
&\left.\bigg(
    R_0\frac{\partial u}{\partial r} + h_0(u - 1)
       \bigg)
  \right|_{r=R_0}=0\label{e:1sinkcav3}
\end{align}
\end{subequations}
where $u(r)=c(r)/c_B$ and  $h_0 = D_{\rm out}/D_{\rm in}$ is a parameter 
gauging the permeability of the  
internal boundary of the spherical cavity.  Here we assume that $D_{\rm out},D_{\rm in}$
are the ligand diffusion coefficient outside and inside the cavity, respectively.
The solution to the problem~(\ref{e:1sinkcav}) is straightforward, and the capture rate by (total 
flux into) the sink yields
\begin{equation}
\label{e:1sinkcavk}
\frac{k}{k_{S_1}} = \frac{h_0}{\epsilon + h_0(1-\epsilon)}
\end{equation}
where $\epsilon = R_1/R_0$. We see that $k \to k_{S_1} = 4\pi D_{\rm in} c_B$ 
in the limit of infinite cavity $R_0\to\infty$. For a finite cavity with a fixed ligand concentration 
outside,~\eref{e:1sinkcavk} has a simple 
interpretation: the rate of capture is enhanced for $h_0>1$, that is, when $D_{\rm out}> D_{\rm in}$.
In the limit of infinitely absorbing boundary (or conversely, infinitely viscous interior), 
the rate of capture is enhanced by a factor $1/(1-\epsilon)$. This becomes very large 
as the sink approaches the inner surface of the cavity. \\
\indent This simple result may have interesting 
implications for the diffusion of ligands within the bacterial periplasm.
This region, comprised between the outer cell membrane and  an inner (cytoplasmic) membrane, 
can be as wide as 40 \% of the total volume in gram-negative bacteria 
and is typically a very shallow layer in gram-positive bacteria. The periplasm is filled with a thick gel-like,
highly crowded matrix~\cite{McNulty:2006aa} and is lined up with 
many arrays of receptors on the inner cytoplasmic membrane,
facing the outer membrane (interior of the cavity).
Many ligands, such as those related to chemotaxis, diffuse to receptors within the inner 
membrane (at $r=R_1$). Since typically $(R_0-R_1)/R_0\ll 1$ and the periplasm is very 
crowded~\cite{McNulty:2006aa}, one has  $D_{\rm in} \ll D_{\rm out}$ and $1-\epsilon \ll 1$, 
which would thence boost the rate of capture.\\
\indent Using the general addition theorems for solid harmonics 
(see details reported in appendix~\ref{a1}),
we are now in a position to answer many interesting questions related to such problems. 

\subsubsection{Two spheres inside a spherical cavity\label{s:5}}

It is interesting to investigate diffusion interactions between two sinks in a finite domain.
Let us consider the simple case of two identical perfect sinks arranged symmetrically along a diameter 
of a spherical cavity with respect to the center. Let us denote with $\ell$ the center-center 
distance, with $R_1$ the size of the sinks and with $R_0$ the size of the cavity, whose internal 
surface is made perfectly absorbing. The problem~(\ref{sistema_def}) can be solved as described in 
appendix~\ref{a1}. The results are summarized in~\fref{f:2sphcav}. In agreement with what discussed
in the previous section, one can appreciate that the rate to the two confined sinks is larger than 
in the absence of cavity. In particular the rate increases abruptly as the sinks approach the 
inner boundary of the cavity. This is a direct consequence of the 
assumption that the ligand density at the cavity interface is equal to the bulk density. Another non-trivial 
observation  is that the normalized rate now depends on the size of the sink: large sinks have more 
capture power with respect to the open-domain, non-confined setting than small ones. Concerning the rate 
of capture of single confined sinks, one remarks that the prediction~(\ref{e:1sinkcavk}) in the limit 
$h_0\to\infty$ for a sink at the center of the cavity is still accurate when the sink is displaced up 
to a distance of $\simeq7\div 8 R_1$ from the center (constant curves with squares in~\fref{f:2sphcav}, left panel).\\
\indent The rates increase in a cavity and diffusive interactions decrease. This effect is illustrated
in the right panel of~\fref{f:2sphcav}. The larger the embedded sinks, the  greater the overall rate
and correspondingly the  weaker the diffusive interactions. For example, for $R_1/R_0=0.2$, the DIs
are practically gone ($k_2 = 2k_1$) already for $\ell \simeq 5\div6 R_1$, i.e. when 
the outer surface of the sinks is at a distance of about $0.2 R_0$ from the inner surface of the cavity.

%
\begin{figure}[t!]
\centerline{\includegraphics[width=12cm]{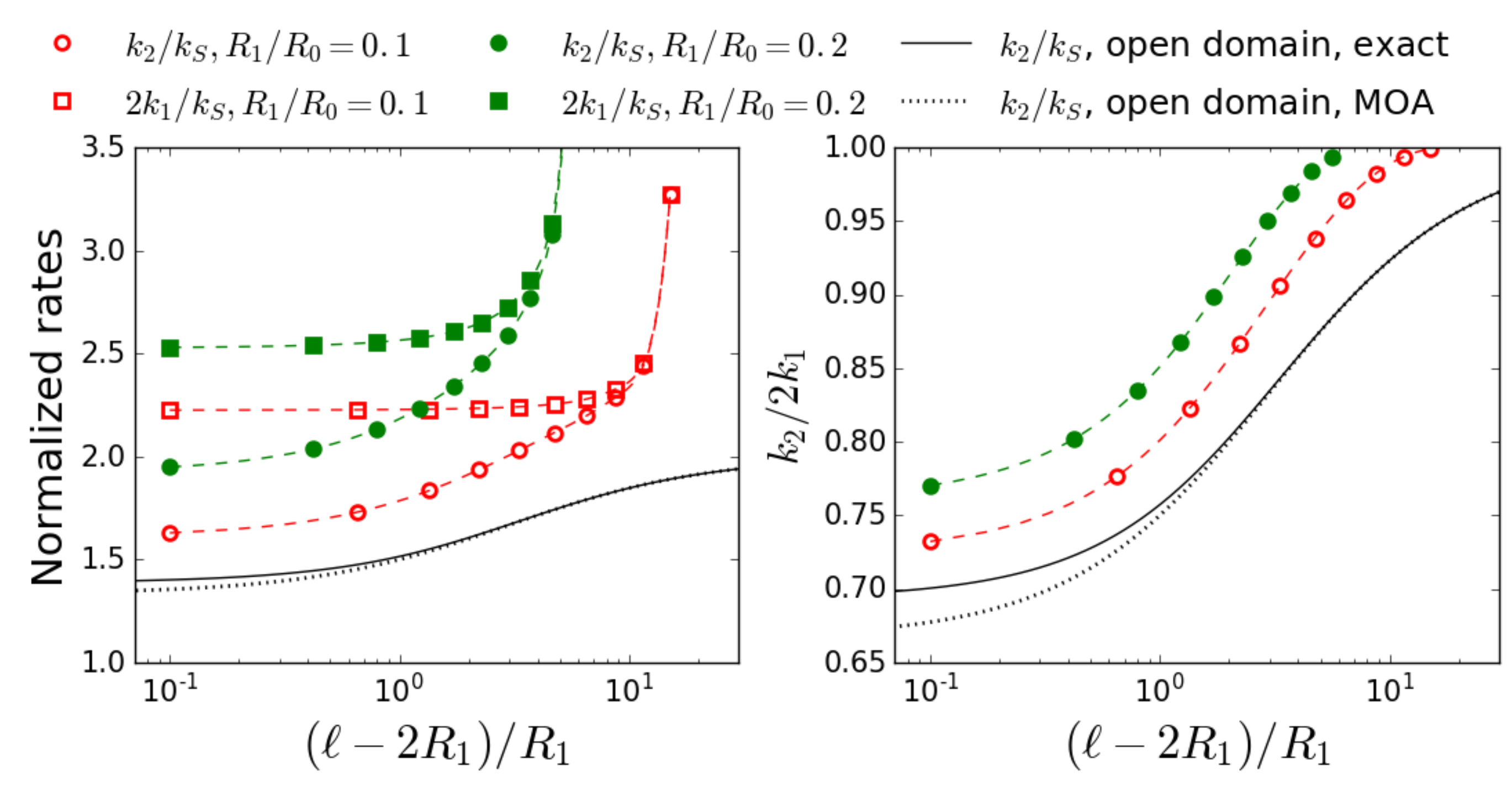}}
\caption{Rate of capture by two identical spheres of radius $R_1$ inside a spherical cavity 
of radius $R_0$ with absorbing inner boundary ($h_0\to\infty$). With reference to a frame with the 
origin at the center of $\Omega_0$, the two spheres are placed symmetrically along the 
$z$ axis at a center-center separation $\ell\in[2R_1,2(R_0-R_1)]$. 
Left: normalized rates of the two-sphere system compared
with twice the rate of one isolated sink at the same position vs rescaled center-center distance.
The rates are normalized to the rate of one isolated sink in the open domain, $k_S = 4\pi D R_1 c_B$.
Right: measure of diffusive interactions. The caption refers to the left panel. Symbols and lines 
in the right panel refer to the same cases as in the left panel.}
\label{f:2sphcav}
\end{figure}
%

\subsubsection{Many sinks on a spherical inner layer inside a spherical cavity\label{s:5}}

%
\begin{figure}[t!]
\centerline{\includegraphics[width=9cm]{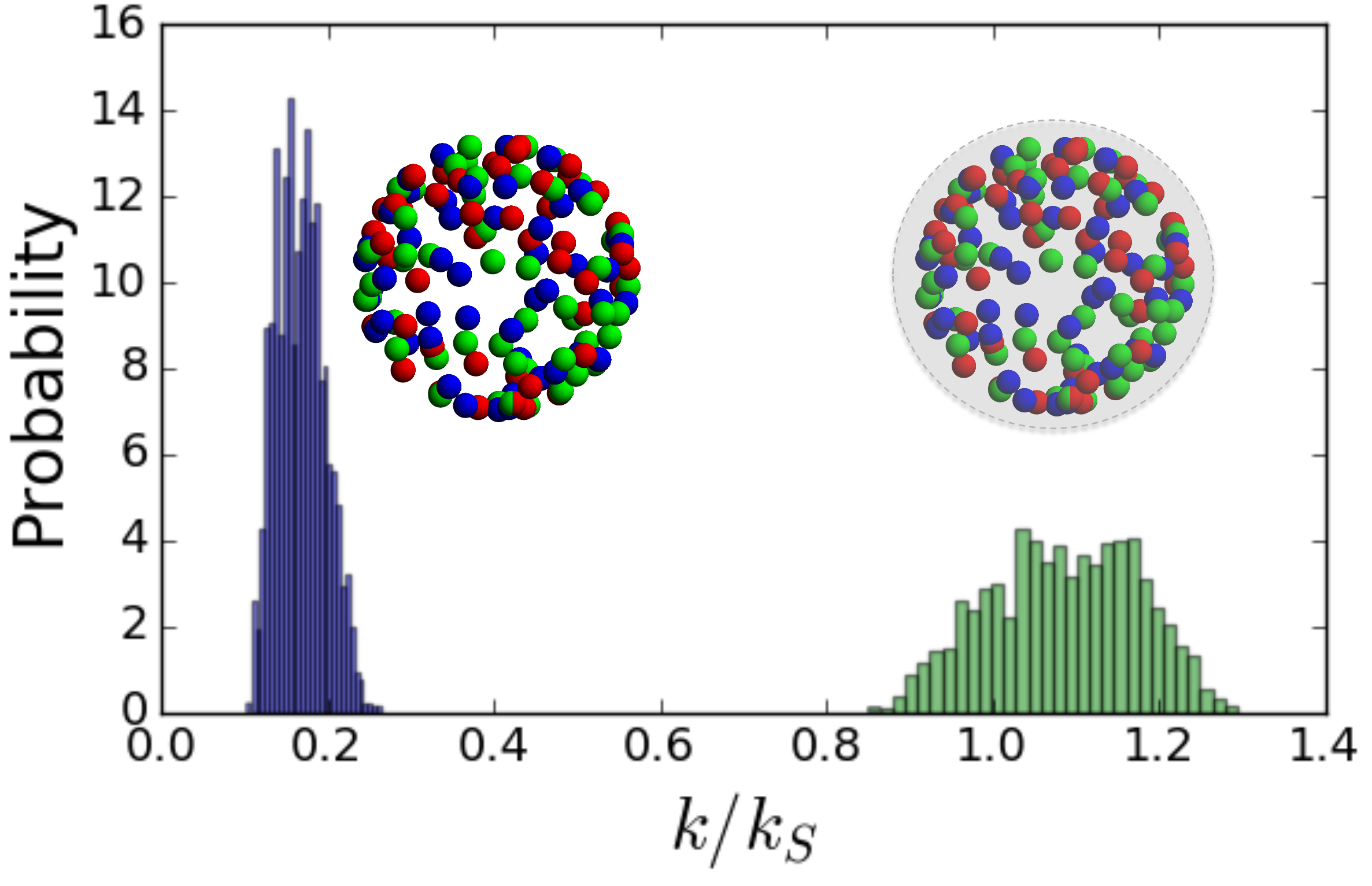}}
\caption{Histograms of the individual capture rate of $N=50$ identical sinks of radius $R_1 = R_0/10$
arranged at a fixed distance $d = 0.95\times (R_0 - R_1)$ from the center of a spherical cavity 
with absorbing inner wall (right) and in the infinite domain (left). 
The histograms are normalized to the capture rate of an isolated 
sink in the open domain, $k_S = 4\pi D R_1 c_B$. The histograms refer  to a population 
of 100 independent random configurations of the $N=50$ sinks, 3 of which are pictured explicitly 
to illustrate the geometry of the problem.}
\label{f:3confallk}
\end{figure}
%

It is instructive to use the above described method to investigate 
the rate of capture of many equivalent sinks arranged randomly 
at a given distance from the center on a spherical layer. In the open domain, the rate to such
ensembles of sinks is strongly reduced due to diffusive interactions. For example, the average capture
rate of random configurations 
of $N=50$ non-overlapping sinks of size $R_1$ at a distance $d=8.85 R_1$ from the center is 
$k_{50} = (8.33 \pm 0.01) k_S$, with $k_S = 4\pi D R_1 c_B$ (average over 100 independent 
configurations). DIs reduce by a staggering 85 \% the overall capture rate of the ensemble 
with respect to as many isolated sinks. We have learned that confining sinks within a cavity 
helps sustain the capture rate due to the proximity (exterior of the cavity) of the 
bulk concentration (effectively reducing the ligand depletion region). In fact, the same
ensembles of sinks within a cavity of radius $R_0 = 10 R_1$, i.e. close to the inner surface
of the cavity, display a rate of capture 
$k^{\rm c}_{50} = (54.2 \pm 0.2) k_S$. This corresponds to a situation of even {\em positive}
cooperativity. This situation is found for example in the bacterial periplasm.
It is reasonable to assume that ligands, whose concentration is constant 
outside the cell, diffuse very slowly in the periplasm as compared to the bulk, which justifies 
the assumption $h_0 \gg 1$. It is fascinating to think that such a complex, double-membrane architecture 
could be a an evolutive answer to the requirement of maximizing the diffusive flux of (possibly low-concentration)
ligands to a set of membrane-bound receptors.\\
\indent\Fref{f:3confallk} reveals what happens to the individual capture rates for a large 
set of equivalent configurations of receptors on the inner membrane of an imaginary periplasmatic 
layer. Each receptor-sink is seen to capture on average the same amount of flux it would capture 
if it was isolated at the center of the cavity (see~\eref{e:1sinkcavk} for $h_0\to\infty$),
i.e. about $1/(1-R_1/R_0) \approx 1.11$ in units of $k_S$.
This somewhat surprising fact is due to the close proximity of the sinks to the inner surface
of the cavity (see also again~\fref{f:2sphcav}). If the cavity disappears, this figure drops down 
to about $ 0.15 \,k_S$ (left histogram in~\fref{f:3confallk}). 
This is another manifestation of the virtual suppression of diffusive 
interactions for sinks close to the absorbing inner surface of a cavity.  
Furthermore, it can be observed that the intrinsic variability of the capture rate around the 
ensemble average is reduced when diffusive interactions are strong (width of the left
histogram in~\fref{f:3confallk}). This means that when DIs are weaker, not only the ensemble 
recovers a large rate of capture on average, but some of the receptors-sink individually 
can attain  large peaks of capture rate.

\section{Summary}

In this short, mostly pedagogical review, we have described the phenomenon of diffusion 
to capture, which has important implications in a wide range of fields in biology and
physical chemistry. We have shown how, under certain circumstances, the problem of bimolecular
encounters and reactions can be solved as a two-body stationary diffusion boundary problem.
This theoretical framework immediately leads to some surprising conclusions. One of the most 
striking findings concerns the rate of ligand capture by a receptor-covered cell. The classic
mean-field solution of this problem shows that a fraction of surface coverage as low as $10^{-4}$
(approximately $10^4$ receptors of $1.5$ nm size on the surface of a cell of size $10$ $\mu$m) 
ensures that the overall rate of capture is of the same order (reduced by a factor of 2) as for a
fully covered surface. Moreover, we have shown that if the same number of 
active receptors are all moved into an active cluster covering the same surface fraction, the 
overall rate of capture drops by a factor of up to $10^2$. This is a first manifestation 
of {\em diffusive interactions}, which describe the interference among diffusive fluxes to 
neighboring reactive boundaries. \\
\indent In order to provide a rigorous mathematical description of diffusive interactions, 
we have considered in detail the classic problem of diffusion to two neighboring sinks at 
a center-to-center distance $\ell$ in the
open domain. Although this problem can be solved by using bispherical 
coordinates~\cite{Samson:1977aa}, we have followed another, more general approach, 
based on translational addition theorems for spherical harmonics~\cite{Morse:1953le}.
The exact solution of the problem, expressed in the form of an infinite series of multipoles,
can be surprisingly well approximated for perfectly absorbing spheres by the monopole term alone,
showing that diffusive interactions are long-range, i.e. decrease as $\ell^{-1}$. \\
\indent The mathematical strategy based on addition theorems can be easily extended to 
compute the rate of capture of an ensemble of spheres of arbitrary, size, intrinsic reactivity 
($\kappa^\ast$) and arranged in arbitrary configurations in 3D, both in the open domain and 
within a spherical cavity. This theory, developed  in Ref.~\refcite{Galanti:2016ab}, 
is described in detail in appendix~\ref{a1}.
Although the applications of such theoretical framework are countless, we have examined here 
two simple examples. First we have studied the case of two sinks within a cavity, whose 
solution shows that diffusive interactions are generally reduced in a finite domain with 
an absorbing inner surface, concomitantly with the enhancement of the rate of capture. This 
phenomenon is due to the fact that, in this modeling strategy, the density of ligands reaches 
its constant bulk value outside the cavity (whose surface is modeled as a permeable membrane),
which enhances the rate of capture of a given boundary with respect to the open domain. 
Interestingly, now the relative position of the boundary within the cavity obviously makes 
a difference, the rate of capture increasing massively as the boundary approaches the 
inner surface of the cavity. At the same time, if many sinks are present, the diffusive interactions
among them are virtually suppressed for many-body configurations close to the inner surface of the cavity.
The second and final example studied, namely many independent configurations of sinks close 
to the inner surface of the cavity, shows this clearly.  Finally, we have argued that this problem, 
while interesting on purely theoretical grounds, might also have important implications in 
ligand-receptor interactions in biology. Notably, ligand diffusion to receptors on the cytoplasmatic 
membrane in the periplasmatic space in bacteria provides an example of this problem. 
In this specific case, ligand diffusion in the crowded, gel-like periplasm is likely to be strongly 
reduced with respect to the mobility in the bulk outside, which justifies modeling the inner 
surface of the outer (cell) membrane as an absorbing boundary. The fascinating speculation that follows 
from these results is that such a complex architecture might  have been designed by evolution 
to maximize the ligand-receptor binding rate. This would make sense, as such receptors are mostly
chemotaxis receptors, used by bacteria to sense gradients of nutrients (small molecules).

\section*{Acknowledgements}
This work was partially supported within the framework of the state task 
program of the FASO Russia (Theme 0082-2014-008, No AAAA-A17-117040310008-5).


\begin{appendix}[Rules for selecting the appropriate addition theorem]
\label{a1}

The addition theorems for spherical harmonics allow one to express 
a combination of spherical harmonics, written in multiple coordinate systems, 
as a function of any one of them. 
Depending on the type of spherical harmonic that one needs to re-expand (regular or irregular) 
and on the geometry of the domain, one among three addition theorems has to be chosen in each specific case.
Let us suppose to have spherical harmonics $u^+(\boldsymbol{r}_\beta)$ and $u^-(\boldsymbol{r}_\beta)$ 
written in a spherical coordinate system centered on $S_\beta$, 
that we want to express at a given point $P$ as a function of the $S_\alpha$-coordinate system 
(see~\fref{f:schemacoord}). The relation $\boldsymbol{r}_\beta= \bold{L}_{\beta\alpha}+\bold{r}_\alpha$ holds.
The regular harmonics $u^+(\bold{r}_\beta)$ are always expressed as a function of the 
regular harmonics  $u^+(\bold{r}_\alpha)$, namely
\begin{equation}
\label{r0-r0}
\begin{split}
&r_\beta^nY_{mn}(\bold{r}_\beta)=\\
&\sum_{q=0}^{n}\sum_{g=-q}^{q} \frac{(n+m)!}{(n-q+m-g)!(q+g)!}L_{\beta\alpha}^{n-q}Y_{m-g,n-q}(\boldsymbol{L}_{\beta\alpha})r_\alpha^{q}Y_{gq}(\boldsymbol{r}_{\alpha}).
\end{split}
\end{equation}
%
%
\begin{figure}[t!]
\centerline{\includegraphics[width=9cm]{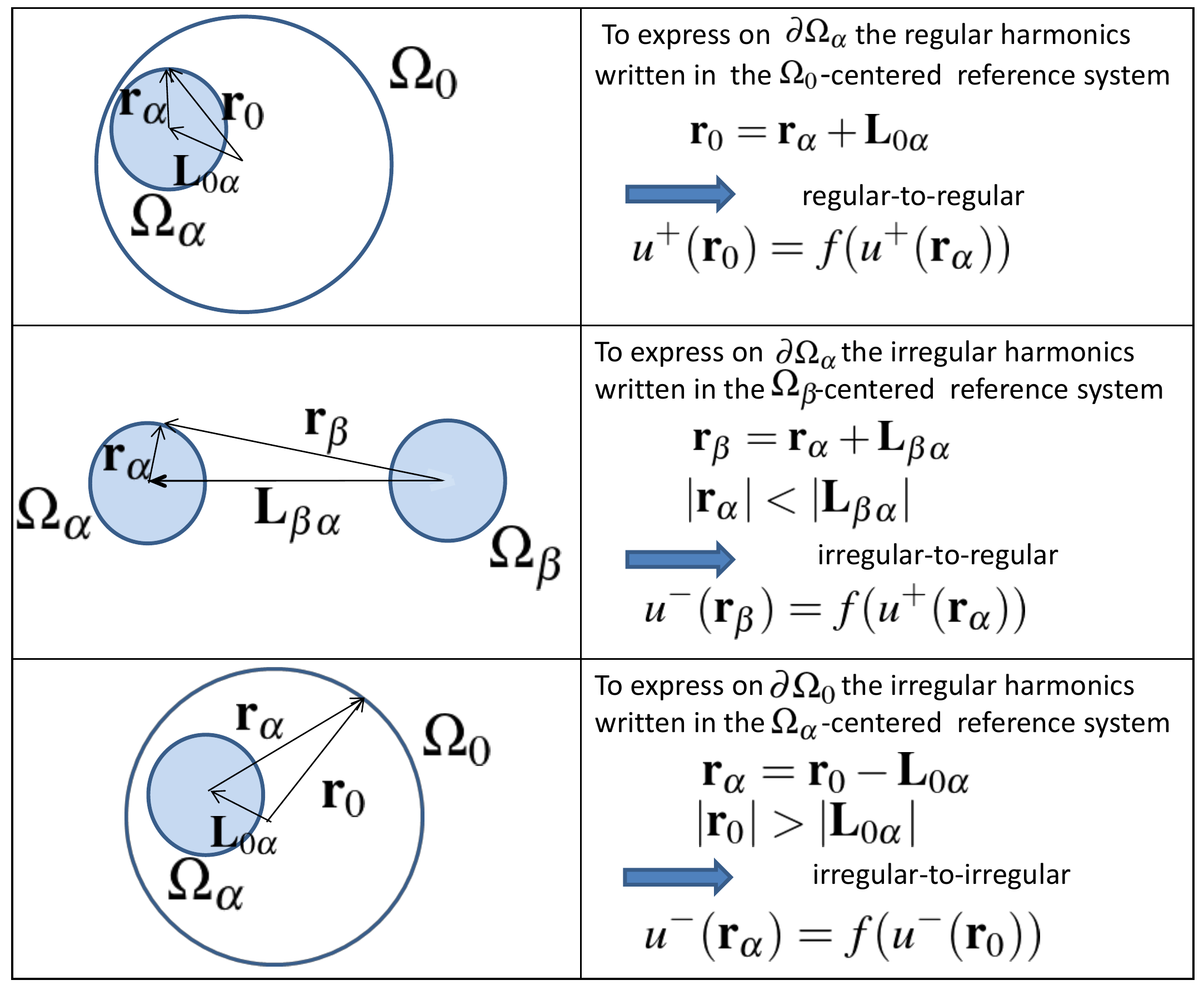}}
\caption{Scheme for the the application of the addition theorems~(\ref{r0-r0}),~(\ref{i0-r0})
and~~(\ref{i0-i0}) to express the boundary conditions in the local coordinates on 
$\partial\Omega_0$ and on each $\partial\Omega_\alpha$. 
The choice of the appropriate addition theorem depends on the ratio 
between the distance between the centers of each pair of reference systems 
and the norm of the position vector in the new reference system.
\label{addi}}
\end{figure}
%
If one has to re-expand an irregular harmonic  $u^-(\boldsymbol{r}_\beta)$, two cases are possible, 
depending on the ratio between the distance $L_{\beta\alpha}$ between the centers of the 
old  and new reference frames, and the norm of the the vector $\boldsymbol{r}_\alpha$ expressing 
the position of $P$ in the new frame $S_\alpha$ .
More precisely, if $|\boldsymbol{r}_{\alpha}|<|\boldsymbol{L}_{\beta\alpha}|$, 
then one has to write the irregular harmonic as a function of the 
regular harmonics centered on $S_{\alpha}$, namely
\begin{equation}
\label{i0-r0}
\begin{split}
&\small r_{\beta}^{-n-1}Y_{mn}(\boldsymbol{r_{\beta}})= \\
&\sum_{q=0}^{\infty}\sum_{g=-q}^{q} (-1)^{q+g}\frac{(n-m+q+g)!}{(n-m)!(q+g)!}
L_{\beta\alpha}^{-(n+q)-1}Y_{m-g,n+q}(\boldsymbol{L}_{\beta\alpha})r_{\alpha}^{q}Y_{gq}(\boldsymbol{r}_{\alpha}).
\end{split}
\end{equation}
Conversely, if  $|\boldsymbol{r}_{\alpha}|>|\boldsymbol{L}_{\beta\alpha}|$, 
then one has to write the irregular harmonic as a function of the irregular 
harmonics centered in $S_{\alpha}$:
\begin{equation}
\label{i0-i0}
\begin{split}
&r_{\beta}^{-n-1}Y_{mn}(\boldsymbol{r_\beta})=\\
&\sum_{l=0}^{\infty}\sum_{s=-n}^{n} \frac{(-1)^{l+s}(n+l-m+s)!}{(n-m)!(l+s)!}L_{\alpha\beta}^l Y_{sl}(-\boldsymbol{L}_{\alpha\beta})r_\alpha^{-(n+l)-1} Y_{m-s,n+l}(\boldsymbol{r_\alpha}). 
\end{split}
\end{equation}
To summarize, one can use the following scheme to change variables 
from system $S_\beta$ to $S_\alpha$ (see also~\fref{addi})
\begin{itemize}
\item $u^+(\boldsymbol{r}_\beta)=f(u^+(\boldsymbol{r}_{\alpha}))$
\item $u^-(\boldsymbol{r}_\beta)=\begin{cases}
f(u^+(\boldsymbol{r}_{\alpha}))\quad\text{if}\quad
|\boldsymbol{r}_{\alpha}|<|\bold{L}_{\beta\alpha}|\\
f(u^-(\boldsymbol{r}_{\alpha}))\quad\text{if}\quad
|\boldsymbol{r}_{\alpha}|>|\bold{L}_{\beta\alpha}|
\end{cases}$.
\end{itemize}

%
\subsection{The solution to the problem}

By using the above addition theorems, the solution to the 
problem~(\ref{sistema_def}) can be cast in the form of the following 
infinite-dimensional linear system,
\begin{equation}
\label{sistemaN}
\displaystyle
\begin{cases}
\displaystyle -B_{gq}^{\alpha}+
\sum_{n=0}^{\infty}\sum_{m=-n}^{n}\bigg(
 A_{mn} H_{m,n}^{(\alpha,g,q)}\mathbf 1 _{q\leq n}+\sum_{\beta=1,\beta\neq 
  \alpha}^{N}B_{mn}^{\beta} W_{m,n}^{(\alpha,\beta,g,q)} \bigg)=0\\
\displaystyle A_{gq}+\sum_{\alpha=1}^{N}\sum_{n=0}^{q}\sum_{m=-n}^{n}B_{mn}^{\alpha}
V_{g,q}^{\alpha,m,n}\mathbf 1 _{\{g-(q-n)\leq m \leq g+(q-n)\}}=\delta_{(g,q)=(0,0)}.\\
\end{cases}
\end{equation}
where
\begin{equation}
\begin{split}
 V_{g,q}^{\alpha,m,n}=-\frac{h_0+q+1}{(q-h_0)}
\frac{(-1)^{q-n+m-g}(q-g)!}{(n-m)!(q-n+m-g)!}\eta_{0\alpha}^{q-n}\chi_{\alpha}^{n+1} Y_{m-g,q-n}(-\boldsymbol{L_{0\alpha}})
\end{split}
\end{equation}
\begin{equation}H_{m,n}^{(\alpha,g,q)}= \frac{(q-h_\alpha)}{(h_\alpha+q+1)}
\binom{n+m}{q+g}  \chi_{\alpha}^q \eta_{0\alpha}^{n-q}Y_{m-g,n-q}(\boldsymbol{L_{0\alpha}})
\end{equation}
\begin{equation}
\begin{split}
W_{m,n}^{(\alpha,\beta,g,q)} =\frac{(q-h_\alpha)}{(h_\alpha+q+1)}&(-1)^{q+g}\times \\
&\frac{(n-m+q+g)!}{(n-m)!(q+g)!}\eta_{\beta\alpha}^{-(n+q)-1} \chi_{\alpha}^{q}
\chi_{\beta}^{n+1}Y_{m-g,n+q}(\boldsymbol{L_{\beta\alpha}})
\end{split}
\end{equation}
with $\displaystyle \chi_\alpha:=\frac{R_\alpha}{R_0}$ and 
$\displaystyle \eta_{\alpha\beta}:=\frac{L_{\alpha\beta}}{R_0}$ with $\eta_{\alpha\beta}=\eta_{\beta\alpha}$.
The system~(\ref{sistemaN}) can be cast in matrix form as follows:
\[ 
\left[
\begin{BMAT}(e)[7pt,5cm,5cm]{c.cccc}{c.cccc}
\1 & V^1 &V^2 &\dots &V^{N} \\
H^1 &- \1& W^{1,2}&\dots&W^{1,N}\\
H^2 & W^{2,1}&-\1&\dots & W^{2,N}\\
\vdots &\vdots&\vdots&\ddots&\vdots\\
 H^{N} &W^{N,1}& W^{N,2} &\dots &-\1\\
 \end{BMAT}
 \right]
  \times 
  \left[
   \begin{BMAT}(r)[2.5pt,0pt,5cm]{c}{ccc.ccc.c.ccc}
A_{00}\\
\vdots\\
A_{N_M N_M}\\
B^{1}_{00}\\  
 \vdots\\
B^{1}_{N_M N_M}\\
\vdots\\
B^{N}_{00}\\  
 \vdots\\
B^{N}_{N_M N_M}
\end{BMAT}
\right]
   = 
\left[
\begin{BMAT}(r)[2.5pt,0pt,5cm]{c}{ccc.ccc.c.ccc}
1\\
\vdots\\
0\\
0\\  
 \vdots\\
0\\
\vdots\\
0\\  
 \vdots\\
0
\end{BMAT}
\right]
\]
%
\subsection{The rate}

The capture rate for a ligand with diffusion coefficient $D$ 
by a selected boundary $\Omega_{\alpha}$ can be computed easily
as the total incoming flux, namely
\begin{equation}
\label{e:rate_alpha_def}
k_\alpha = -\int_{\partial\Omega_\alpha} \boldsymbol{J}_\alpha \cdot \hat{\boldsymbol{n}} \, dS 
\end{equation}
where $ \boldsymbol{J}_\alpha = -D \nabla_\alpha c$ is the current to the 
$\alpha$-th boundary. It is not difficult to see 
from the general form of the solution~(\ref{e:gensolu}) and general 
properties of the Legendre polynomials $P^m_n(\mu_\alpha)$ that~\eref{e:rate_alpha_def} gives
\begin{equation}
\label{e:rate_alpha}
\frac{k_\alpha}{k_{S_\alpha}} = -B_{00}^\alpha
\end{equation}
where $k_{S_\alpha} = 4\pi D R_\alpha c_B$ is the Smoluchowski rate of capture corresponding 
to an isolated sink of radius $R_\alpha$ in the infinite domain.

\end{appendix}

%
%
%


\end{document}